\documentclass[aps,prl,floatfix,twocolumn,superscriptaddress]{revtex4}
\usepackage{amsmath,amssymb}
\usepackage{graphicx}
\usepackage{epsfig}
\usepackage[usenames]{color}

\begin{document}

\title{Radiation Pressure Quantization}

\author{V.~M.~Kovalev}
\affiliation{A.V. Rzhanov Institute of Semiconductor Physics, Siberian Branch of Russian Academy of Sciences, Novosibirsk 630090, Russia}
\affiliation{Department of Applied and Theoretical Physics, Novosibirsk State Technical University, Novosibirsk 630073, Russia}


\author{A.~E.~Miroshnichenko}
\affiliation{School of Engineering and Information Technology, University of New South Wales, Canberra, ACT 2600, Australia}

\author{I.~G.~Savenko}
\affiliation{Center for Theoretical Physics of Complex Systems, Institute for Basic Science, Daejeon, Republic of Korea}

\date{\today}
\begin{abstract}
Kepler's observation of comets tails initiated the research on the radiation pressure of celestial objects and 250 years later they found new incarnation after the Maxwell's equations were formulated to describe a plethora of light-matter coupling phenomena. Further, quantum mechanics gave birth to the photon drag 
effect. 
Here, we predict a universal phenomenon which can be referred to as {\em quantization of the radiation pressure}. We develop a microscopic theory of this effect which can be applied to a general system containing Bose--Einstein-condensed particles, which possess an internal structure of quantum states. By analyzing the response of the system to an external electromagnetic field we find that such drag results in a flux of particles constituting both the condensate and the excited states. We show that in the presence of the condensed phase, the response of the system becomes quantized which manifests itself in a {\em step-like} behavior of the particle flux as a function of electromagnetic field frequency with the elementary quantum determined by the internal energy structure of the particles.
\end{abstract}

\maketitle



\textit{Introduction.---} Ponderomotive force of light acting on atoms, molecules and other particles results in a momentum transfer between light and matter~\cite{Gibson, Ivchenko} --- the phenomenon referred to as the radiation pressure. Historically, the hypothesis of radiation pressure was for the first time suggested by J.~Kepler in the beginning of XVII century in an attempt to explain why the tails of comets point back from the Sun. In frameworks of classical electrodynamics, the radiation pressure was considered by J.~Maxwell in 1870th. It was shown to be closely connected with light scattering and absorption by particles.

In frameworks of quantum mechanics, radiation pressure is a result of momentum transfer from a photon to a system, for instance, an atom~\cite{Werij} or a molecule. In condensed matter, light pressure results in a current of charge carriers and it is called the photon drag effect (PDE). The first theory of this phenomenon was based on electron--photon interaction mediated by an acoustic phonon~\cite{Grinberg2, Yee}. Charge carriers, such as free electrons and holes, can absorb radiation by means of interaction with an electromagnetic field (EMF), and they are forced to move in a direction of the wave vector of light. PDE has been extensively studied in semiconductors~\cite{Costa, Berman, Shalygin}, dielectrics~\cite{Loudon}, metals~\cite{Goff, Kurosawa}, monolayer and multilayer graphene~\cite{Glazov, Entin}, carbon nanotubes~\cite{Mikheev}, topological insulators~\cite{Plank}, 2D electron gas~\cite{Wieck, Luryi, Grinberg}, and other systems.

According to classical description of the PDE in semiconductors, the drag current reads
$\textbf{j}(\omega)\sim \textbf{k}\alpha(\omega)I$,
where $\textbf{k}$ is the wave vector, $I=cE^2/8\pi$ is the intensity of the electromagnetic wave and $\alpha(\omega)$  is the  coefficient of light absorption by the charge carriers. Evidently, the frequency dependence of the drag current is determined by the spectrum of the absorption coefficient. In the majority of cases, this dependence is monotonous or resonant if the frequency of the EMF $\omega$ is close to the energy of quantum transitions in the system. However, it is not a general rule.

\begin{figure*}[t]
\includegraphics[width=0.9\linewidth]{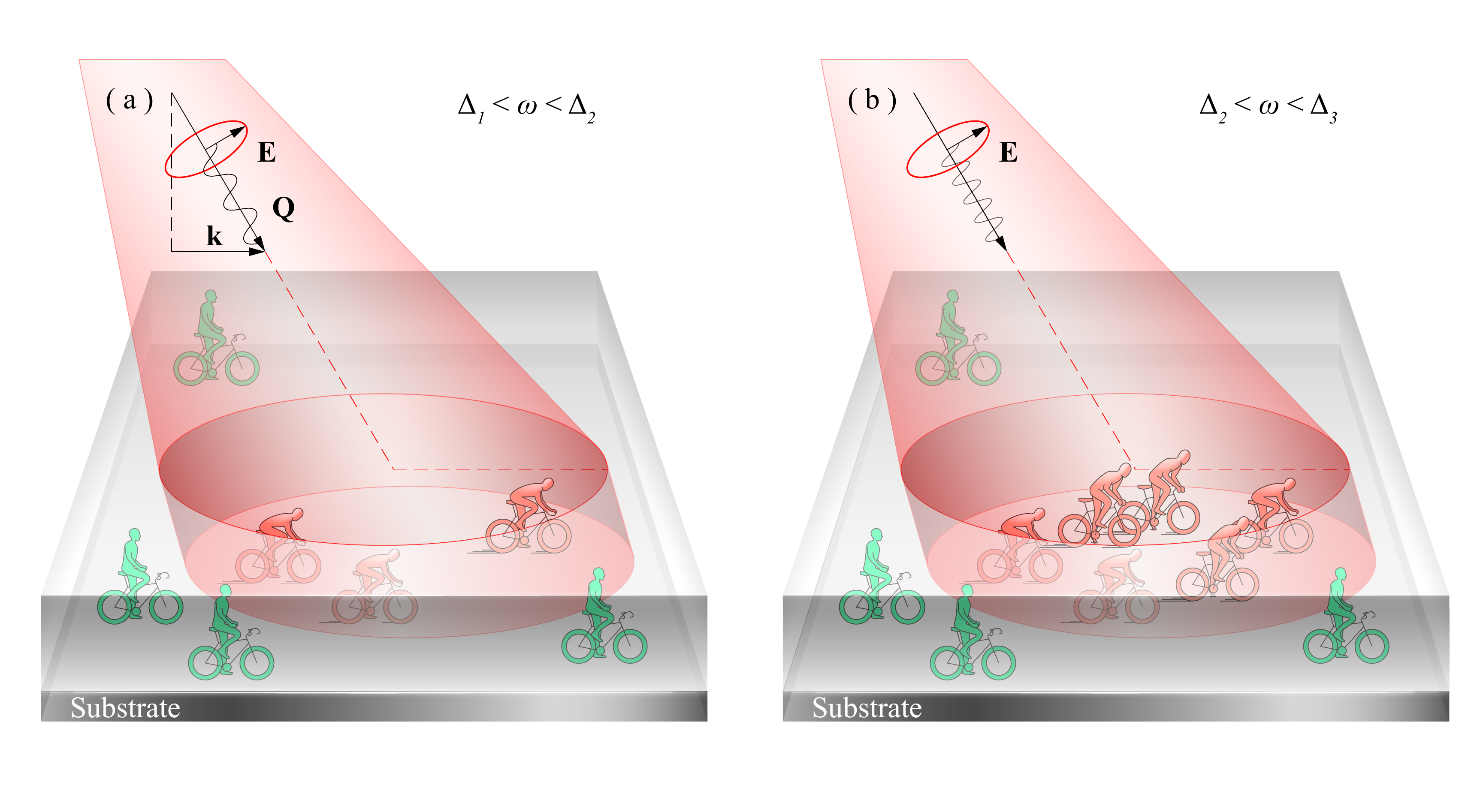}
	\caption{System schematic. Bosons (presented as bicycles) are exposed to an electromagnetic field. It results in their flux in the direction collinear with the in-plane projection $\textbf{k}$ of the wave vector of light $\textbf{Q}$. Current after the second threshold $\omega=\Delta_2$ (a) is bigger than the current at the first threshold $\omega=\Delta_1$ (b) since more bosons take part in the current.}
\label{Fig1}
\end{figure*}

In this Letter, we study the effect of radiation pressure on a purely quantum system of bosons containing particles in condensed quantum state. We will show that in a Bose--Einstein condensate (BEC), the drag current of bosons and thus the response of the system become quantized. It is a universal phenomenon which can be possibly observed in atomic and solid state condensates, thus we will consider a general model of a Bose gas, in which each boson possesses an internal structure of quantum states, which is essential for the theory developed below. The spectrum of a single boson with an eigenfunction $|\eta,\textbf{p}\rangle$ reads
\begin{eqnarray}\label{equation2}
\varepsilon_\eta(\textbf{p})=\varepsilon(\textbf{p})+\Delta_\eta,
\end{eqnarray}
where $\varepsilon(\textbf{p})=\textbf{p}^2/2M$ is a kinetic energy of the particle center-of-mass motion, and $\Delta_\eta$ is the energy spectrum of the internal motion. It can be a spectrum of an atom (in a cold atomic condensate system) or the energy of relative motion of an electron and a hole constituting an exciton (in excitonic BECs). Here index $\eta$ stands for the whole set of quantum numbers which characterize the internal spectrum of the particle and the value $\eta=0$ refers to the lowest energy state (ground state) of the internal spectrum of Bose particles and all the energies will be measured from $\Delta_{\eta=0}$. We will assume that before being irradiated, the system is in the Bose-Einstein condensed state $|\eta,\textbf{p}\rangle$ where all the particles are in the ground $\eta=0$ state of their internal motion, zero kinetic energy $\textbf{p}=0$ of their center-of-mass motion without a dipole moment. 

\emph{Model.---}
Let us consider a system of bosons exposed to an EMF with (Fig.~1) the wavelength exceeding the size of a particle thus allowing us to use a dipole approximation.
The electric field then depends on the center-of-mass coordinate $\textbf{r}$ only,
$\textbf{E}(x)=\textbf{E}_0e^{i\textbf{kr}-i\omega t}+\textbf{E}^*_0e^{-i\textbf{kr}+i\omega t}$,
and the light--matter coupling can be described by the matrix elements $\textbf{d}_{21}\cdot\textbf{E}$.
Here the indices $1,~2$ stand for the ground and excited quantum states of the internal particle motion, $|1\rangle\equiv|\eta=0\rangle$ and $|2\rangle\equiv|\eta\neq0\rangle$. Then $\textbf{d}_{12}=\langle1|\textbf{d}|2\rangle$ is a matrix element of dipole moment operator of the particle. For simplicity, we assume that initially the particles do not possess a dipole moment, $\textbf{d}_{11}=\textbf{d}_{22}=0$, and $\varepsilon_1(\textbf{p})=\varepsilon(\textbf{p})$, $\varepsilon_2(\textbf{p})=\varepsilon(\textbf{p})+\Delta_\eta$ are the energies of the ground and excited states, correspondingly. 

The system response to a pressure of external EMF is a current of particles which is determined by the light absorption coefficient. Qualitatively, the BEC--EMF interaction Hamiltonian reads
\begin{eqnarray}\label{equation2.1}
\textbf{d}_{2 1}\cdot\textbf{E}_0\sum_{\textbf{p}}c^\dag_{\eta,\textbf{p}+\textbf{k}}(t)a_{\textbf{k}}(t)c_{0,\textbf{p}}(t),
\end{eqnarray}
where $c_{\eta,\textbf{p}}(t)=c_{\eta,\textbf{p}}(0)\exp(-i\varepsilon_\eta(\textbf{p})t)$ and $a_{\textbf{k}}(t)=a_{\textbf{k}}(0)\exp(-i\omega t)$ are the annihilation operators for the Bose particle and EMF photon, respectively. The theoretical description of BEC is based on the Bogoliubov theory of a weakly interacting Bose gas. It requires that the Bose gas is dilute, $na^d\ll1$, where $n$ is the particle concentration and $a$ is a particle size and $d$ is system dimensionality. In order to describe the dynamics of the BEC, we will use a Gross-Pitaevskii equation (GPE). In its framework, low-energy excitations of the BEC represent Bogoliubov quasiparticles (bogolons) with the dispersion $\omega_\textbf{p}=\sqrt{\varepsilon_\textbf{p}(\varepsilon_\textbf{p}+2gn_c)}=sp\sqrt{1+(p\xi)^2}$, where $s=\sqrt{gn_c/M}$, $\xi=1/(2Ms)$ are the sound velocity and the healing length, $g$ is interparticle interaction strength, $n_c$ is density of particles in the BEC. In a long-wavelength limit $\xi p\ll1$ (that is equivalent to $\varepsilon_\textbf{p}\ll gn_c$) the dispersion law of the bogolons becomes linear, $\omega_\textbf{p}=sp$. We will consider $T=0$ thus disregarding the processes of thermal excitation of bogolons. 
Further we present $c_{0,\textbf{p}}$ in the form
\begin{eqnarray}\label{equation2.2}
c_{0,\textbf{p}}(t)=c_{0,0}\delta(\textbf{p})+u_\textbf{p}b_\textbf{p}(t)+v_\textbf{p}b^\dag_{-\textbf{p}}(t),
\end{eqnarray}
where $c_{0,0}$ describes the particles in the BEC state with zero momentum and $|c_{0,0}|^2=n_c$. Here $u_\textbf{p}$ and $v_\textbf{p}$ are the Bogoliubov transformation coefficients and $b_\textbf{p}(t)=b_\textbf{p}(0)\exp(-i\omega_\textbf{p}t)$ are Bogoliubov excitation operators. Substituting Eq.~(\ref{equation2.2}) into Eq.~(\ref{equation2.1}), we can come up with several principal quantum channels of the EMF absorption.

\begin{figure}[t]
\includegraphics[width=0.6\linewidth]{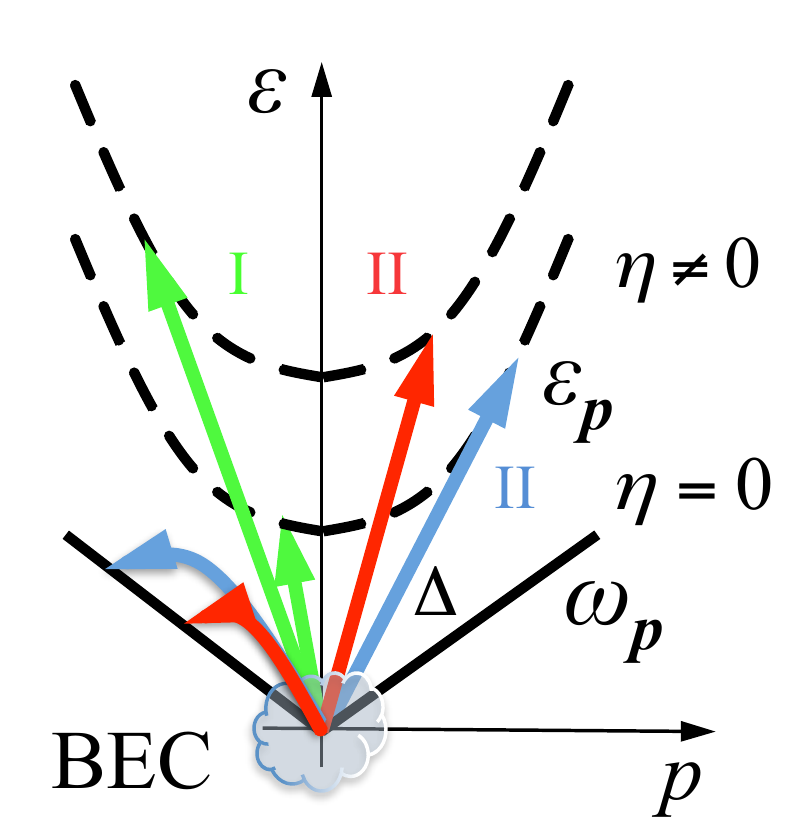}
	\caption{Excitation spectrum of the internal degrees of freedom of the boson during absorption of a quantum of EMF.}
\label{Fig2}
\end{figure}

\textit{Results and discussion.---} The first term in Eq.~(\ref{equation2.2}) describes a transition of a Bose particle from the BEC to an excited state with $\eta\neq 0$ with the energy conservation law $\omega=\varepsilon_\textbf{k}+\Delta_\eta\approx\Delta_\eta$. Such processes give a set of resonances which correspond to the number of levels of internal particle motion, $\Delta_{\eta\neq 0}$, see Fig.~\ref{Fig2} transitions I.   

Beside these, there exists another type of transitions described by the second and third terms in Eq.~(\ref{equation2.2}). They can be referred to as the Belyaev processes~\cite{Chaplikkovalev, OurArXiveBelyaev} and happen when the light absorption is accompanied by not only excitation of the particle but also the emission or absorption of a Bogoliubov excitations of the condensate, see Fig.~\ref{Fig2} transitions II. The corresponding energy conservation law reads $\omega=\varepsilon_{\textbf{p}+\textbf{k}}+\Delta_\eta+\omega_{-\textbf{p}}$, where $\omega_\textbf{p}$ is the bogolon dispersion. Further analysis shows that such processes result in the quantization of the BEC responce.

Indeed, the probability of the absorption of a photon is proportional to $\sum_\textbf{p}v^2_\textbf{p}\delta(\omega-\varepsilon_{\textbf{p}-\textbf{k}}-\Delta_\eta-\omega_{-\textbf{p}})$. In the long-wavelength limit, $\omega_\textbf{p}\approx sp$, and accounting for the fact that $sp\gg\varepsilon_{\textbf{p}+\textbf{k}}$ and $v^2_\textbf{p}\sim\omega^{-1}_\textbf{p}$, in a 2D case we find the thresholdlike behavior:
\begin{eqnarray}\label{equation19}
\sum_\eta\Bigl|\textbf{d}_{\eta 1}\cdot\textbf{E}_0\Bigr|^2\int\limits_0^\infty \frac{pdp}{\omega_\textbf{p}}\delta(\omega-\Delta_\eta-\omega_{-\textbf{p}})\sim\\\nonumber\sim\sum_\eta\Bigl|\textbf{d}_{\eta 1}\cdot\textbf{E}_0\Bigr|^2\theta[\omega-\Delta_\eta].
\end{eqnarray}
Thus taking into account the internal structure of the particles leads to the quantization of the response of the system to an external light pressure.


%
%
\begin{figure}[b!]
\includegraphics[width=0.9\linewidth]{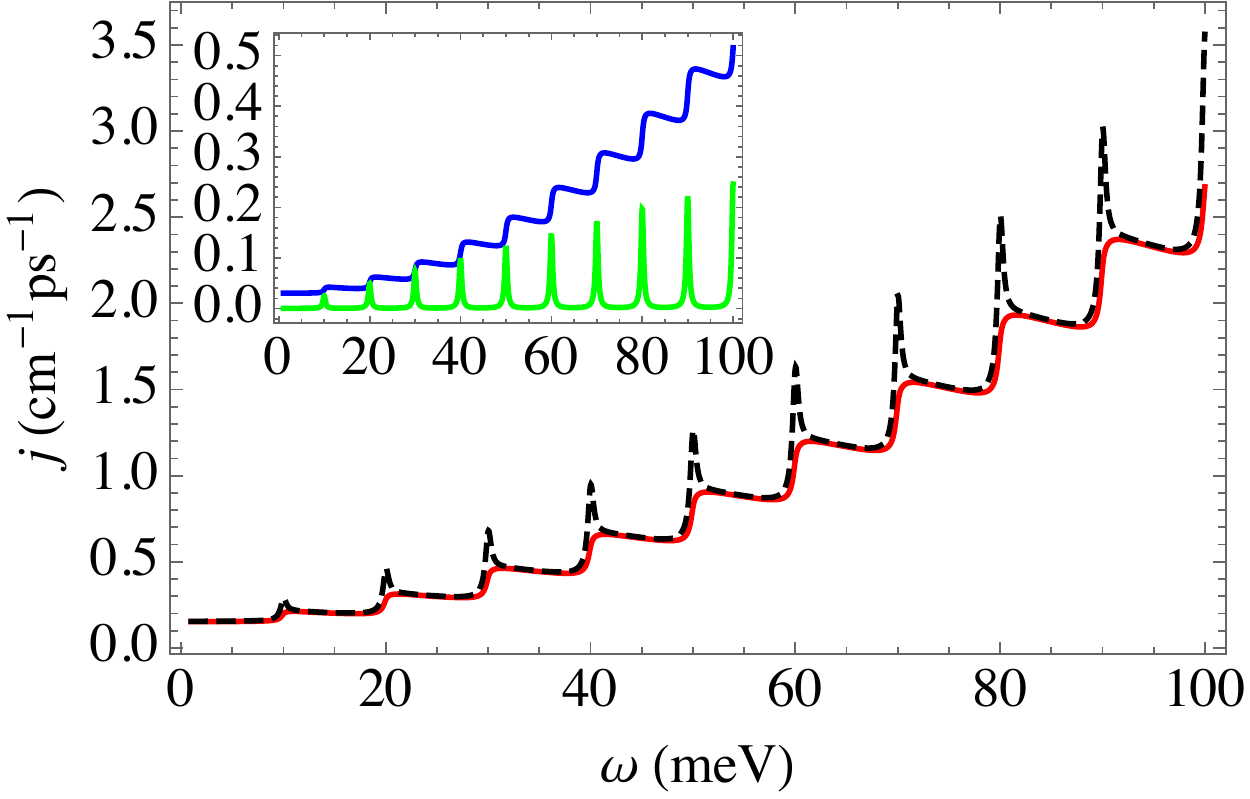}
	\caption{Spectrum of the total current density $j_{c}=j_{c1}+j_{c2}$ (main plot). Two components of the current $j_{c1}$ (green curve) and $j_{c2}$ (blue curve) according to Eqs.~(\ref{equation17}) and~(\ref{equation18}) (inset) for the parameters $\Delta=10$ meV, $M=0.5\cdot m_0$, $n_c=2\cdot10^{13}$ cm$^{-2}$ (red curve), $n_c=2\cdot10^{14}$ cm$^{-2}$ (black dashed curve), $n_c=5\cdot10^{13}$ cm$^{-2}$ (Inset).}
\label{Fig3}
\end{figure}

If the bosons are in normal phase, their current spectrum represents a set of resonances (see Supplemental Material~\cite{SM}). In the presence of the BEC, total drag current consists of two components (see~\cite{SM} for details): 
\begin{eqnarray}\label{equation17}
\textbf{j}_{c1}&=&
\frac{2n_c\textbf{k}\tau^2}{M\hbar}
\sum_\eta
\Bigl|\textbf{d}_{1\eta}\cdot\textbf{E}_0\Bigr|^2
\\
\nonumber
&&\times
\left[\frac{1}{1+4\tau^2(\omega-\Delta_\eta)^2}
-
\frac{1}{1+4\tau^2(\omega+\Delta_\eta)^2}\right],
\end{eqnarray}
and
\begin{eqnarray}\label{equation18}
\textbf{j}_{c2}&=&\frac{5\tau \textbf{k}}{8\pi\hbar^2}
\sum_\eta
\Bigl|\textbf{d}_{1\eta}\cdot\textbf{E}_0\Bigr|^2
\\
\nonumber
&&\times
\left(\arctan\Bigl[2\tau(\omega+\Delta_\eta)\Bigr]+\arctan\Bigl[2\tau(\omega-\Delta_\eta)\Bigr]\right),
\end{eqnarray}
where $\tau$ is particle  lifetime in the excited states, which we take independent of $\eta$. 
Evidently, both the components~(\ref{equation17}) and~(\ref{equation18})  share some similar properties, they are i) collinear with the momentum of the EMF (since $\sim\textbf{k}$) and ii) proportional to the intensity of the EMF, $I\sim |\textbf{E}_0|^2$. 

Figure~\ref{Fig3} shows the spectrum of the current density and its components for the parameters taken for bosons in solid-sate. We assumed that $\Delta_\eta$ is equidistant, $\Delta_\eta=\eta\cdot\Delta$. In general, it is not the case and one has to consider the selection rules for the internal transitions between quantum states.
The transition matrix element was taken to be $|\textbf{d}_{12}\cdot\textbf{E}_0|=0.01\Delta$. %
Obviously, this value is controlled by the amplitude of the external EMF obeying the condition $|\textbf{d}_{12}\cdot\textbf{E}_0|\ll\Delta$ (since the perturbation theory is only applicable if the external light is reasonably weak). 
The most interesting are the terms in the second lines in both the Eqs.~(\ref{equation17}) and~(\ref{equation18}) and the sums over the states $\eta$. Evidently, these terms are proportional to the absorption coefficient. While Eq.~(\ref{equation17}) has resonant behavior, Eq.~(\ref{equation18}) obeys step-like behavior (see Fig.~\ref{Fig3}, Inset). Summing up~(\ref{equation17}) and~(\ref{equation18}), we find the total current in the system, $\textbf{j}_c=\textbf{j}_{c1}+\textbf{j}_{c2}$ (see Fig.~\ref{Fig3}, main plot).

\emph{Conclusions.---} 
We predicted a universal phenomenon of radiation pressure quantization and developed a microscopic theory of this effect, applied to a general system containing Bose--Einstein-condensed particles. 
We found that under the pressure of an external electromagnetic field there appears a drag flux of particles constituting both the condensate and the excited states. Moreover, in the presence of the condensed phase, this current becomes quantized.

We would like to stress several factors. First of all, the theory presented here is universal. It can be applied to any Bose condensates which possess internal degrees of freedom. Most of bosons such as cold atoms, excitons, exciton-polaritons possess this property.

Second, quantization of the response of a BEC to external radiation pressure can manifest itself in a number of other effects, in which the processes of light absorption play major role, such as Raman scattering (since the scattering cross-section is proportional to the imaginary part of the linear response function), acoustic and acousto-electric effects~\cite{Parmenter}, acoustic drag in condensates of hybrid particles, detection of ``dark'' condensates~\cite{Butov}.

The authors would like to thank A. Chaplik and M. Entin for fruitful discussions. We acknowledge the support by the Russian Science Foundation (Project No.~17-12-01039), the Institute for Basic Science in Korea (Project No.~IBS-R024-D1), and the University of New South Wales Scientia Fellowship. 


%
%





%
%


\begin{thebibliography}{50}

\bibitem{Gibson} A. F. Gibson, M. F. Kimmitt, and A. C. Walker, Photon drag in germanium, Appl. Phys. Lett. \textbf{17}, 75 (1970).


\bibitem{Ivchenko} E. L. Ivchenko, Optical Spectroscopy of Semiconductor Nanostructures (Alpha Science, Harrow, UK, 2005).

\bibitem{Werij} H. G. C. Werij, et al., Light-induced drift velocities in Na-noble-gas mixtures, Phys. Rev. Lett. \textbf{58}, 2660 (1987).

\bibitem{Grinberg2} A. A. Grinberg, Zh. Eksperim. i Teor. Fiz. \textbf{58} 989 (1970) [Sov. Phys. JETP \textbf{31}, 531 (1970)].

\bibitem{Yee} J. H. Yee, Theory of Photon-Drag Effect in Polar Crystals, Phys. Rev. B \textbf{6}, 2279 (1972).

\bibitem{Costa} C. Rodrigues-Costa and O. A. C. Nunes, Theory of photon-drag effect in bulk magnetic semiconductors, Phys. Rev. B \textbf{46}, 15046 (1992).

\bibitem{Shalygin} V. A. Shalygin, M. D. Moldavskaya, S. N. Danilov, I. I. Farbshtein, and L. E. Golub, Circular photon drag effect in bulk tellurium, Phys. Rev. B \textbf{93}, 045207 (2016).

\bibitem{Berman} O. L. Berman, R. Ya. Kezerashvili, and Yu. E. Lozovik, Drag effects in a system of electrons and microcavity polaritons, Phys. Rev. B \textbf{82}, 125307 (2010).

\bibitem{Loudon} R. Loudon, S. M. Barnett, and C. Baxter, Radiation pressure and momentum transfer in dielectrics: The photon drag effect, Phys. Rev. A \textbf{71}, 063802 (2005).

\bibitem{Goff} J. E. Goff and W. L. Schaich, Theory of the photon-drag effect in simple metals, Phys. Rev. B \textbf{61}, 10471 (2000).

\bibitem{Kurosawa} H. Kurosawa and T. Ishihara, Opt. Express \textbf{20}, 1561 (2012).


\bibitem{Glazov} M.M. Glazov, S.D. Ganichev, High frequency electric field induced nonlinear effects in graphene, Physics Reports \textbf{535}, 101 (2014).

\bibitem{Entin} M. V. Entin, L. I. Magarill, and D. L. Shepelyansky, Theory of resonant photon drag in monolayer graphene, Phys. Rev. B \textbf{81}, 165441 (2010).

\bibitem{Mikheev} G. M. Mikheev, A. G. Nasibulin, R. G. Zonov, A. Kaskela, and E. I. Kauppinen, Nano Lett. \textbf{12}, 77 (2012).

\bibitem{Plank} H. Plank et al., Photon drag effect in (Bi$_{1-x}$Sb$_x$)$_2$Te$_3$ three-dimensional topological insulators, Phys. Rev. B \textbf{93}, 125434 (2016).


\bibitem{Wieck} A. D. Wieck, H. Sigg, and K. Ploog, Observation of resonant photon drag in a two-dimensional electron gas, Phys. Rev. Lett. \textbf{64}, 463 (1990).

\bibitem{Luryi} S. Luryi, Photon-Drag Effect in Intersubband Absorption by a Two-Dimensional Electron Gas, Phys. Rev. Lett. \textbf{58}, 2263 (1987).

\bibitem{Grinberg} A. A. Grinberg and S. Luryi, Theory of the photon-drag effect in a two-dimensional electron gas, Phys. Rev. B \textbf{38}, 87 (1988).

\bibitem{Chaplikkovalev} M. V. Boev, A. V. Chaplik , V. M. Kovalev, Interaction of Rayleigh waves with 2D dipolar exciton gas: Impact of Bose-Einstein condensation, J. Phys. D: Appl. Phys. \textbf{50}, 484002 (2017).

\bibitem{OurArXiveBelyaev} M. V. Boev, V. M. Kovalev, and I. G. Savenko, Bogolon--mediated electron capture by impurities in hybrid Bos--Fermi systems, arXiv:1802.06228 (2018).

\bibitem{SM} see Supplemental Material for the derivation of formulas for the components of the total current density.


\bibitem{Parmenter} R. H. Parmenter The Acousto-Electric Effect, Phys. Rev. \textbf{89}, 990 (1953).

\bibitem{Butov} L. V. Butov, Exciton condensation in coupled quantum wells, Sol. State Comm. \textbf{127} (2), 89 (2003); and Condensation and pattern formation in cold exciton gases in coupled quantum wells, J. Phys.: Cond. Matt. \textbf{16}, R1577 (2004); and Cold exciton gases in coupled quantum well structures, J. Phys.: Cond. Matt. \textbf{19}, 295202 (2007).

%
%
%





















\end{thebibliography}
\end{document}